\documentclass[conference,a4paper]{IEEEtran}

\usepackage{etex}
\usepackage{mathtools}
\usepackage{float}
\usepackage{url}
\usepackage{amssymb}
\usepackage{amsmath}
\usepackage{algorithm,algorithmicx,algpseudocode}

\usepackage{pictex,graphicx}

\usepackage{cite}

\usepackage{array}

\usepackage[tight,footnotesize]{subfigure}

\newtheorem{remark}{Remark}

\newcommand{\sgn}{\operatorname{sgn}}

\begin{document}

\title{A Deterministic Annealing Approach to Witsenhausen's Counterexample
}


\author{
  \IEEEauthorblockN{Mustafa~Mehmetoglu }
  \IEEEauthorblockA{Dep. of Electrical-Computer  Eng.\\
    UC Santa Barbara, CA, US\\
    Email: mehmetoglu@ece.ucsb.edu} 

      \and
  \IEEEauthorblockN{ Emrah~Akyol}
  \IEEEauthorblockA{Dep. of Electrical  Eng.\\
   USC, CA, US\\\
    Email: eakyol@usc.edu}
      \and
  \IEEEauthorblockN{Kenneth Rose}
  \IEEEauthorblockA{Dep. of Electrical-Computer   Eng.\\
       UC Santa Barbara, CA, US\\\
    Email: rose@ece.ucsb.edu}

}

\maketitle


\begin{abstract}
``To be considered for an IEEE Jack Keil Wolf ISIT Student Paper Award."
This paper proposes an optimization method, based on information theoretic ideas, to a class of distributed control problems. As a particular test case, the well-known and numerically ``over-mined" problem of decentralized control and implicit communication, commonly referred to as Witsenhausen's counterexample, is considered. The key idea is to randomize the zero-delay mappings. which become ``soft", probabilistic mappings to be optimized in a deterministic annealing process, by incorporating a Shannon entropy constraint in the problem formulation. The entropy of the mapping is controlled and gradually lowered to zero to obtain deterministic mappings, while avoiding poor local minima. For the particular test case, our approach obtains new mappings that shed light on the structure of the optimal solution, as well as achieving a small improvement in total cost over the state of the art in numerical approaches to this problem. Proposed method is general and applicable to any problem of similar nature.
\end{abstract}

\section{Introduction}
Witsenhausen's counterexample (WCE) \cite{witsenhausen} to a conjecture in decentralized control theory is a deceptively simple stochastic control problem that has remained unsolved for several decades. The fact that the sum of jointly Gaussian random variables is Gaussian yields a simple solution to the centralized linear quadratic Gaussian (LQG) control problems where optimal control mappings are known to be linear in the observations. This observation led to a natural conjecture that linear control mappings remain optimal even in decentralized settings. However, Witsenhausen provided an example of a decentralized LQG control problem wherein non-linear mappings outperform linear mappings. The problem has been viewed as a benchmark in stochastic networked control, see eg. \cite{serdar} for a detailed treatment. 

Beyond its key strategic importance in decentralized control systems, WCE has implications as an ``implicit communication" problem. Connections between information theory and WCE were first studied in \cite{bansal}, and extensions of the original problem to vector settings were studied from an information theory perspective, see eg. \cite{grover, ubli}. 

In this paper, we propose an optimization method, derived from information theoretic principles, which can be applied to a class of distributed control problems, and specifically to WCE. There has been a significant amount of prior work on optimization methods (see e.g.  \cite{deng, baglietto, lee, li, karlsson}), which can be clustered into two camps. The first camp is based on optimization of ``structured" continuous mappings where a parametric form is assumed and these parameters are optimized. The other camp attacks a related discrete problem with the argument that the discrete problem asymptotically approaches to the original analog problem, at increasingly fine discretization. The approach in this paper belongs to the first camp and employs a powerful nonconvex optimization method, namely deterministic annealing (DA), for optimization process.

DA is based on a statistical physics interpretation of information theory (\cite{roserd},\cite{da}), see also \cite{merhav} for an analysis of the  relationship between statistical physics and information theory. DA introduces controlled randomization into the optimization process where the expected cost is minimized subject to a constraint on the level of randomness as measured by the Shannon entropy of the system. The resultant Lagrangian functional and parameter are analogous to the ``free energy" and the ``temperature", respectively, of a corresponding physical system. The optimization is equivalent to an annealing process that starts by minimizing the cost (free energy) at a high temperature. This minimum is then tracked at successively lower temperatures (lower levels of entropy) as the system typically undergoes a sequence of phase transitions through which the complexity of the solution grows. As the temperature approaches zero, the original cost term dominates the free energy and hard (nonrandom) mappings are obtained.

%
  
  This paper is organized as follows. In Section \ref{sec:prelim} we give the problem definition of WCE. In Section \ref{sec:prior}, we review some of the prior results in the literature. Proposed method is described in Section \ref{sec:method} and the experimental results are given in Section \ref{sec:results}. Discussion and concluding remarks are in Section \ref{sec:disc}.
  
\section {Preliminaries and Problem Definition}
\label{sec:prelim}
Let $\mathbb R$, $\mathbb P\{\cdot\}$ and $\mathbb E\{\cdot\}$ denote the set of real numbers, probability and expectation operations, respectively. $\mathbb E\{\cdot|\cdot\}$ is the conditional expectation, $H(\cdot)$ and $H(\cdot|\cdot)$ are the entropy and conditional entropy. $\nabla_{x}f$ denote the partial derivative of $f$ with respect to $x$. Upper case letters are used to denote random variables and lower case letters for their realizations.

\begin{figure}
\centering \includegraphics[width=1 \linewidth]{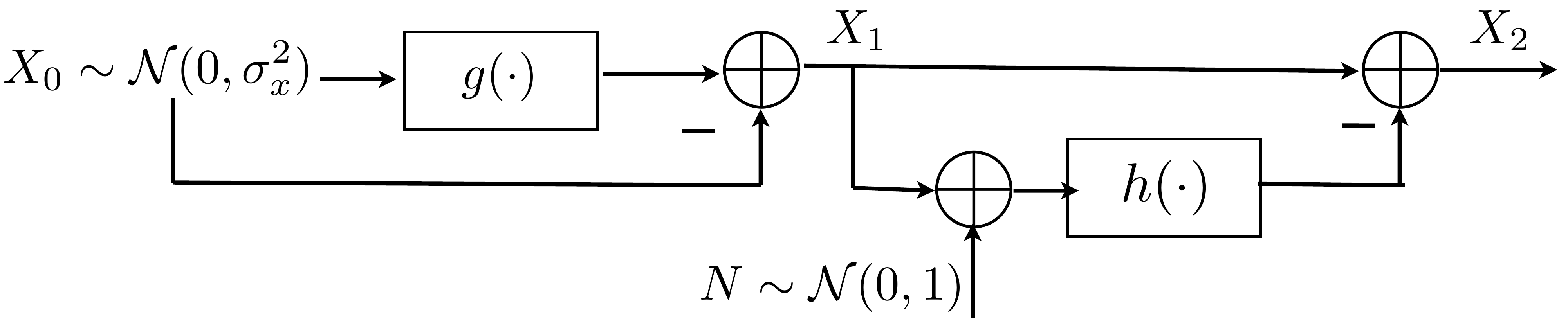}
\caption{The problem setting. Witsenhausen showed a nonlinear $g(\cdot)$ minimizes $ E\{X_2^2\}+k^2  E\{X_1^2\}$}
\label{side} 
\end{figure}

The problem setting is given in Figure \ref{side}. The source $X_0$ and noise $N$ are independent zero-mean Gaussians with variances $\sigma_X^2$ and 1, respectively. The two controllers ${ g}:\mathbb R\rightarrow \mathbb R$ and ${ w} :  \mathbb R  \rightarrow  \mathbb R$ aim to minimize the cost
\begin{equation}
D = \mathbb E \{k^2X_1^2+X_2^2\}
\label{eq:cost}
\end{equation}
where $X_1= X_0+g(X_0)$ and $X_2=X_1-h(X_1+N)$. The given constant $k^2$ is a weight to trade off the control cost $\mathbb E \{X_1^2\}$ with the estimation error $\mathbb E \{X_2^2\}$.

Since the output $X_2$ is the estimation error of $X_1$, the second controller that minimizes the mean squared error distortion is given in closed from as 
\begin{equation}
h(Y)=\mathbb E\{X_1|Y\}.
\label{eq:opt dec}
\end{equation} 
where $Y=X_1+N$. We refer to $g(\cdot)$ as the encoder and to $h(\cdot)$ as the decoder for obvious information theoretic reasons.

\section{Prior Results}
\label{sec:prior}
Since the decoder is in closed form, research on this problem has been focused on finding the optimal encoder $g$. Witsenhausen has derived some properties of $g(\cdot)$, including the property of symmetry about the origin. For a given set of problem parameters ($k=0.2$, $\sigma_X=5$) Witsenhausen provided the following encoder (that outperforms any affine solution): $g(X)=\sigma_X \sgn(X)$ where $\sgn(\cdot)$ is the signum function. This type of solution is referred to as a ``1-step" solution, since the function consists of a single ``step" in the positive half of real line (due to symmetry around the origin, positive half is enough to describe a given solution). Further improvements has been found by various researchers that utilize 2.5, 3, 3.5 and 4-step solutions (an $x.5$ step solution means there is a step that straddles the origin, thus half of it is on the positive portion of real line). Moreover, the latter solutions made improvements by using slightly sloped steps rather than constant ones. Some of the prior results appeared in the literature are given in Table \ref{table:prior results}, including the best result to date.

\begin{table}[H]
\normalsize
\centering
\caption{Major Prior Results}
\label{table:prior results}
\begin{tabular}{rl}
\hline
Solution & Cost \\
\hline
Optimal Affine Solution & 0.961852 \\
1-step, Witsenhausen \cite{witsenhausen} & 0.404253 \\
2-step,\cite{deng} & 0.190 \\
Sloped 2.5 - step, \cite{baglietto} & 0.1701 \\
Sloped 3.5 - step, \cite{lee} & 0.1673132 \\
Sloped 3.5 - step,  \cite{li}& 0.1670790 \\
Sloped 4 - step, \cite{karlsson}& 0.16692462 \\
\hline
\end{tabular}
\end{table}

\section{Proposed Method}
\label{sec:method}
In this section we describe a method based on the concept of deterministic annealing. The cost is optimized by searching for the encoding function $g$ within a parametric class of functions where each function is specified by (a) a partition of space and (b) a parametric local model for each partition. The crucial idea in DA is to introduce randomness into the optimization process, wherein the deterministic partition of space is replaced by a random partition, i.e. every input point is associated with each one of regions in probability. During the minimization of the cost, the Shannon entropy of the system is controlled, gradually lowered, and a deterministic partition (encoder) is obtained at the limit of zero entropy. 

While DA is founded on information theoretic principles, it is motivated by statistical physics and the analogy is emphasized herein to provide further intuition. The entropy-constrained Lagrangian cost is in fact the free energy of a corresponding physical system, with the Lagrange parameter playing the role of ``temperature" in the system. At high temperature, where entropy is maximum, the system effectively has a single local model that dominates the entire space. During the ``annealing" process, where temperature is gradually lowered, the system goes through ``phase transitions" which correspond to increase in the number of local models. At zero temperature, the Lagrangian reduces to the original cost function whose minimum is achieved by a deterministic encoder, thus at this stage the desired solution is obtained. In the rest of this section we give detailed derivation of the proposed method.
 
The parametric encoder functions are specified by local models denoted as $g_m(x)$ and the space partitions $\mathbb R_m$, where $m=\{1,2,...,M_{max}\}$, such that the encoder can be written as 
\begin{equation}
g(x)=g_m(x) \quad \text{for}\,\,\, x \in \mathbb R_m 
\end{equation}
While noting that local models can be in any parametric form, for the particular case of Witsenhausen counterexample we use affine local models given by 
\begin{equation}
g_m(x)=a_mx+b_m.
\end{equation}

 To derive a DA based approach, we introduce randomness into the optimization process by defining association probabilities
\begin{equation}
p(m|x) = \mathbb P\{g(x) = g_m(x)\} = \mathbb P\{x \in \mathbb R_m\}.
\label{eq:prob}
\end {equation} 
for each $x,m$. Accordingly, the system has a Shannon entropy $H(X,M)=H(X)+H(M|X)$. Since the first term is a constant determined by source, we conveniently remove it and define 
\begin{equation}
H \triangleq H(M|X) =  \mathbb E\{\log p(M|X)\}. 
 \label{eq:entropy}
 \end{equation} 
where $H$ measures the average level of uncertainty in the partition of space. 

 \begin{remark}
By employing the so-called mass constrained DA approach \cite{rose1993constrained}, we could equivalently minimize the mutual information $I(M;X)$ instead of maximizing the conditional entropy, obtaining direct relation with rate-distortion theory, see \cite{roserd} for details.
\end{remark}

In order to minimize the cost at a specified level of randomness, we define the Lagrangian 
\begin{equation}
F = D - TH
\label{eq:free}
\end{equation}
referred to as (Helmholtz) ``free energy" and the Lagrange parameter $T$ called ``temperature", to emphasize the intuitively compelling analogy to statistical physics. The algorithm starts at high temperature, where minimization of (\ref{eq:free}) effectively maximizes the entropy. Accordingly, the association probabilities are uniform and all models are identical, or effectively, there is a single distinct local model. As the temperature is decreased, a bifurcation point is reached where the current solution is no longer a minimum, such that there exist a better solution with the local models divided into two or more groups. Intuitively, the current solution becomes a saddle point and a slight perturbation of local models will trigger the discovery of the new solution with increased number of effective local models. Such bifurcations are referred to as ``phase transitions" and the corresponding temperatures are called ``critical temperatures". See \cite{da} for phase transition analysis in the general DA setting. In order to trigger phase transitions, we always keep a duplicate for each local model and perturb them at each temperature. Until the critical temperature is reached, they will merge back during free energy optimization, but will split at phase transitions. At lower values of $T$, randomness is traded for reduction in $J$. In the limit $T = 0$, minimizing $F$ corresponds to minimizing $D$ directly, which produces a deterministic mapping. Therefore, in the practical algorithm we start at a high value of $T$ and gradually lower it while minimizing $F$ at each step.

A brief sketch of the algorithm can be given as follows.
\begin{enumerate}
\item Start at high temperature, single model.
\item Duplicate local models.
\item Minimization of $F$.
\begin{enumerate}
\item Optimize $p(m|x)$ for all $m,x$.
\item Optimize $a_m(x)$ and $b_m(x)$, for all $m$, using gradient descent.
\item Optimize $w(\cdot)$. 
\item Go to Step 4 if $F$ has converged, go to (a) otherwise. 
\end{enumerate}
\item Stop if desired or lower temperature and go to Step 2.
\end{enumerate}

The minimization of free energy in step 3 is done by iteratively optimizing the association probabilities, local model parameters and decoder until the decrease in $F$ becomes insignificant. 

We now derive the optimal association probabilities. Writing the Lagrangian cost in (\ref{eq:free}) as  
\begin{equation}
F = \sum \limits_{m} D_m(x) p(m|x) + T \sum \limits_m p(m|x) \log p(m|x).
\end{equation}
where 
\begin{equation}
D_m(x)=\mathbb E \{k^2X_1^2+X_2^2 | M=m\}.
\end{equation}
Setting $\nabla_{p(m|x)} F=0$, we have 
 \begin{equation}
 D_m(x) + T \log p(m|x) + T = 0
\end{equation}
which yields
\begin{equation}
p(m|x) \propto e^{-[D_m(x)-T]/T}.
\end{equation}
Finally we normalize to obtain
\begin{equation}
p(m|x) = \frac{e^{-[D_m(x)]/T} }{ \sum\limits_{m}e^{-[D_m(x)]/T}} \quad \forall x,m.
\label{eq:opt prob}
\end{equation}

The optimal association probabilities take the form of the Gibbs distribution, which emerge at equilibrium in statistical physics. A fundamental principle of statistical physics is that the minimum of the free energy determines the distribution at thermal equilibrium, at which point the system is governed by the Gibbs distribution. 

Some intuition to the annealing process can be obtained by observing how (\ref{eq:opt prob}) evolves as the temperature is decreased. At the limit of high temperature, (\ref{eq:opt prob}) yields uniform distributions and every input $x$ is equally associated with all $g_m(\cdot)$. As $T$ is lowered, the distributions become more discriminating. As $T \rightarrow 0$, we obtain {\it hard} associations, i.e. every $x$ is fully associated with the local model that contributes least to the distortion (the one with minimum $D_m(x)$).

The local model parameters $a_m(x)$ and $b_m(x)$ can be optimized at any given temperature by any optimization method such as line search or gradient descent. The optimum decoder is given in (\ref{eq:opt dec}).

\section{Experimental Results}
\label{sec:results}

\begin{figure*}
\centering \includegraphics[width=1 \linewidth]{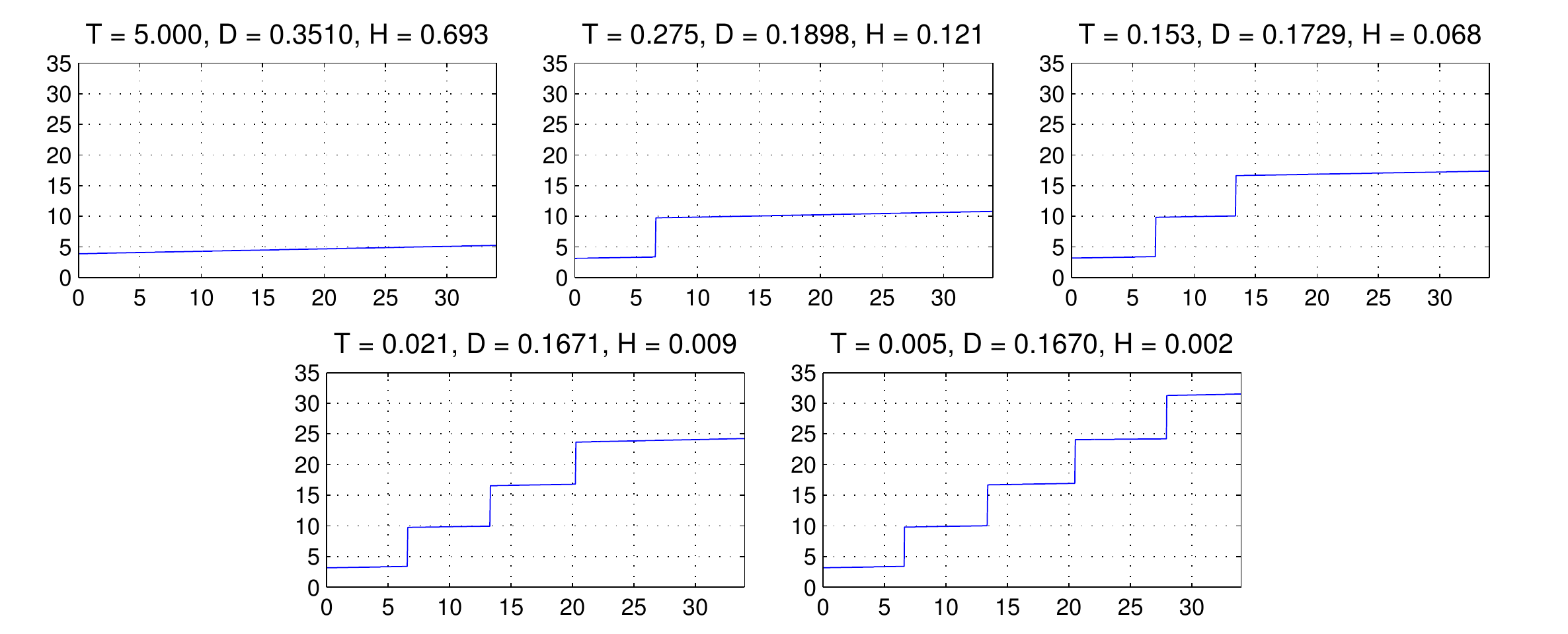}
\caption{Encoder during various phases of the annealing process.}
\label{fig:evol} 
\end{figure*} 

We give the details of the annealing process with its phase transitions in Figure \ref{fig:evol}. At high temperature, there is only one local model, which corresponds to a 1-step solution. As the temperature is lowered, the system undergoes phase transitions that increase the number of local models. Each phase transition reveals a new step for the encoder. One can observe that the phase transitions effectively generate the entire class of $n$-step solutions - an important advantage of the proposed method. In order to generate a solution for a particular $n$, one needs to run the method until the desired number of steps (i.e. local models) is obtained, and then decrease the temperature without growing the model size.

\begin{figure}[H]
\centering \includegraphics[width=1 \linewidth]{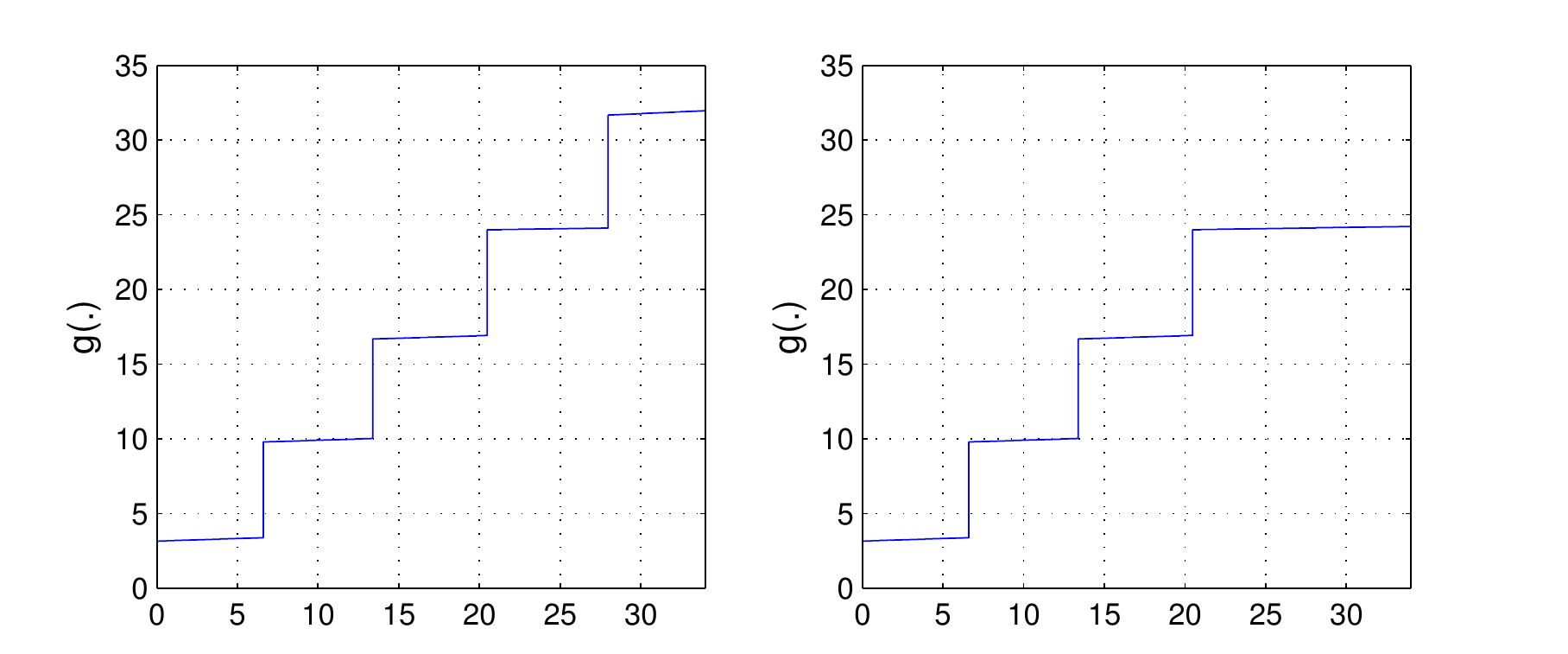}
\caption{5-step and 4-step solutions.}
\label{fig:our encoders} 
\end{figure}

In this work we calculated the solutions up to $n=5$, whereas noting that more steps possibly exist. The cost obtained for 3, 4 and 5-step solutions are given in Table \ref{table:our results}. In addition to the improvement over the previously reported costs, our algorithm revealed a fifth step in the solution. Although the improvement in the cost is very small with this new step, we note the theoretical importance pertaining to the structure of the solution. The 5-step and 4-step encoders are given in Figure \ref{fig:our encoders}.

\begin{table}[H]
\normalsize
\centering
\caption{Our Results}
\label{table:our results}
\begin{tabular}{rl}
\hline
Solution & Cost \\
\hline
3 - step & 0.16694471 \\
4 - step & 0.16692319 \\
5 - step & 0.16692291 \\
\hline
\end{tabular}
\end{table}

\begin{figure}
\centering \includegraphics[width=1 \linewidth]{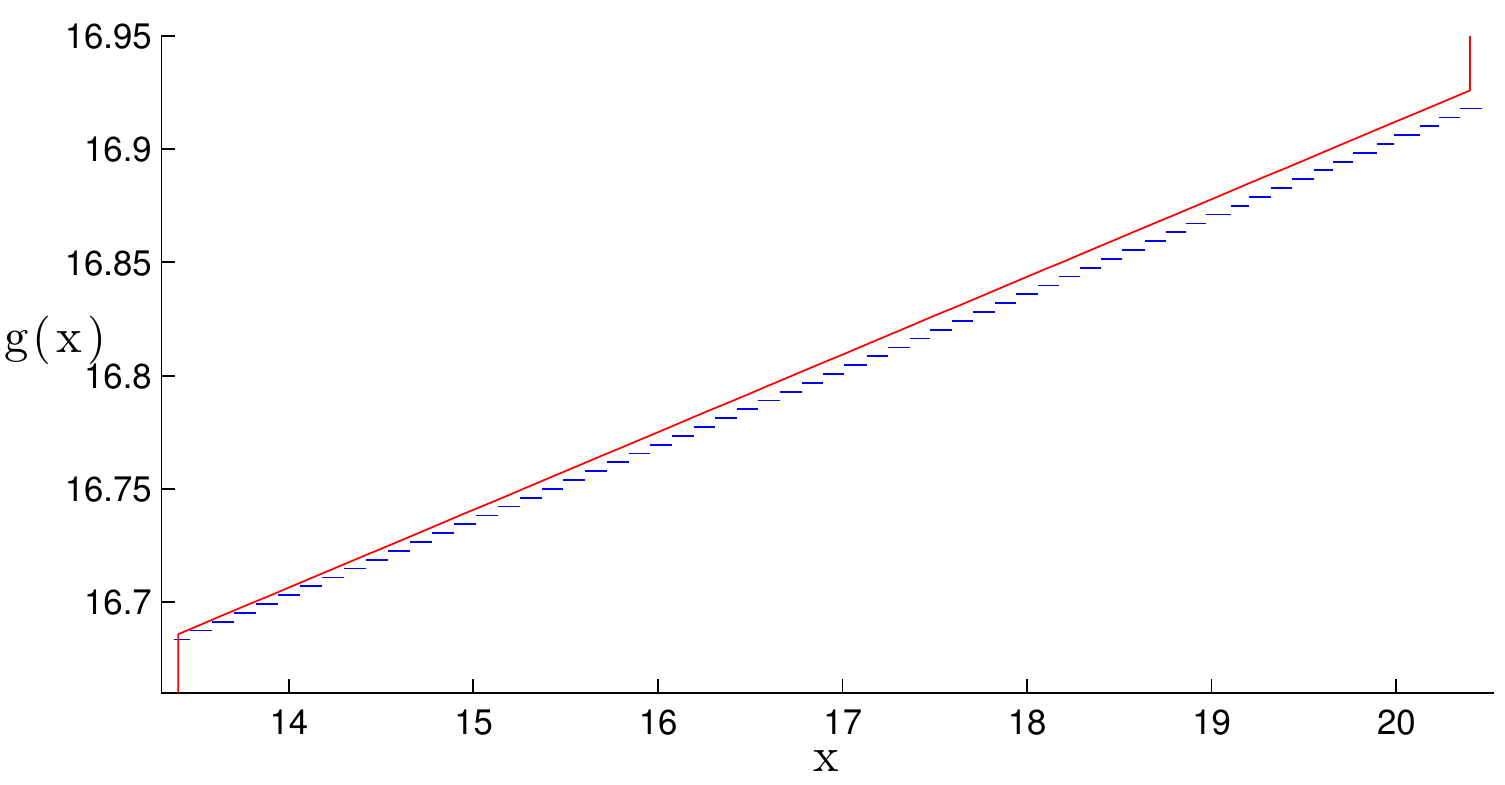}
\caption{Comparison view of the third step in our 4-step solution (straight line) and the 4-step solution in \cite{karlsson}.}
\label{fig:difference 1} 
\end{figure}

\begin{figure}
\centering \includegraphics[width=1 \linewidth]{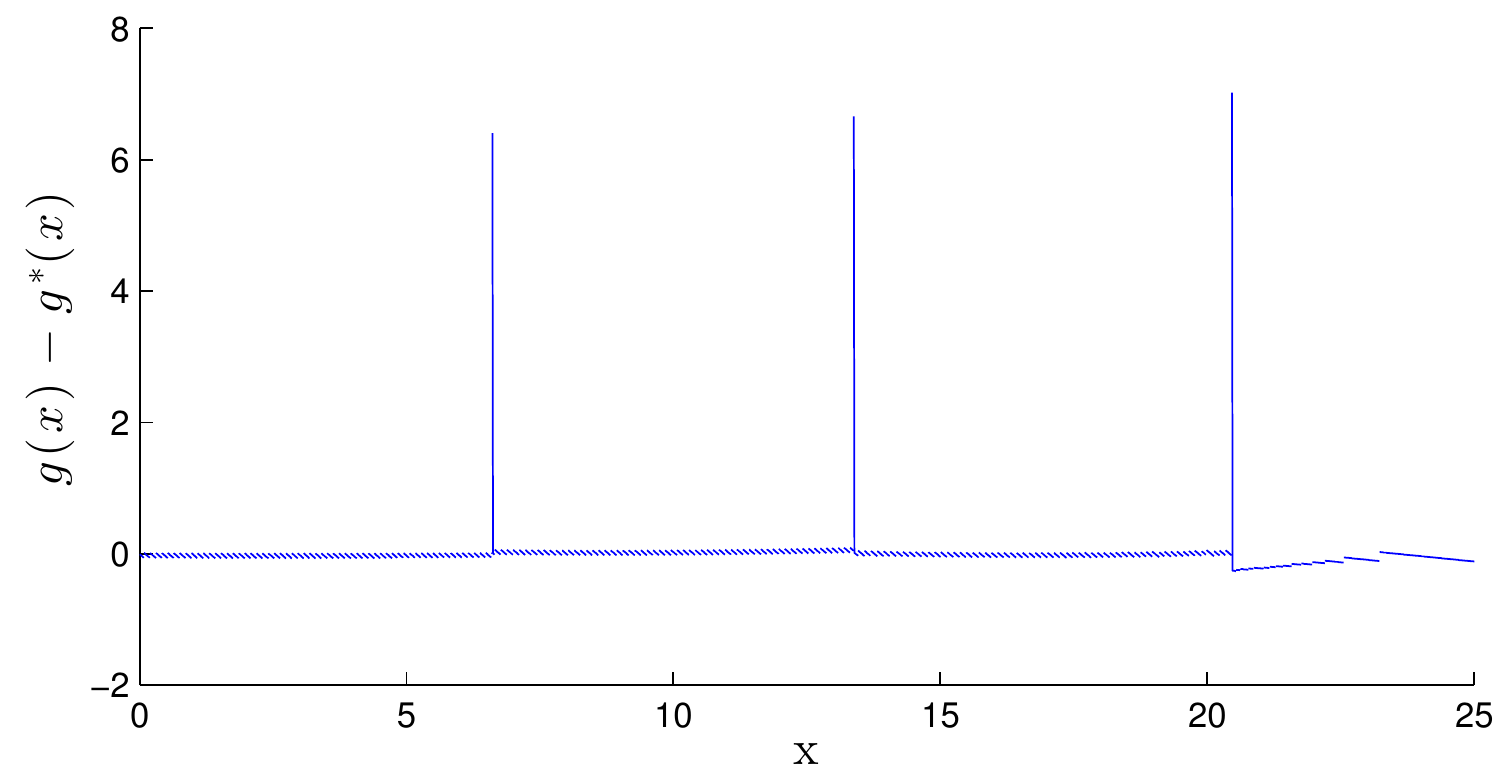}
\caption{The difference between the 5-step solution proposed in this paper ($g(x)$) and the 4-step solution in \cite{karlsson} ($g^*(x)$).}
\label{fig:difference 2} 
\end{figure}

We also compare our results to the best previous result in \cite{karlsson}. The difference between the two encoders is plotted in Figure \ref{fig:difference 2}. There are three main differences: i) Our best solution has 5 steps. ii) Each step is exactly linear as illustrated in Figure \ref{fig:difference 1}. iii) The step boundaries differ slightly as illustrated in Figure \ref{fig:difference 2}.

There are several advantages of the method proposed here. i) We perform optimization process in the original, analog domain, without discretization. This approach yields the linearity of the \textit{steps} as illustrated in Figure \ref{fig:difference 1}. ii) We employ a powerful non-convex optimization tool, DA. \cite{karlsson} uses ``noisy channel relaxation" \cite{gadkari1999robust} (NCR) originally developed for vector quantizer design purposes. NCR offers improvement over greedy techniques thanks to the ad hoc relaxation it employs to avoid local optima, however, it has been outperformed by DA which is derived from basic principles, in a variety of optimization settings, including most relevantly in the optimization of zero-delay source-channel mappings \cite{mehmetoglu_itw}.

\begin{remark}
\label{remark:cost calculation}
The costs here are calculated according to (\ref{eq:cost}). For integrations, we used the same numerical methods as described in \cite{karlsson}.
\end{remark}

{\bf Note}: Matlab code for our calculations of the total cost, including our decision functions can be found in \cite{witsen_webpage}. 

\section {Discussion}
\label{sec:disc}
In this paper, we proposed an optimization method for distributed control problems, whose solutions are known to be non-linear. As an example we showed the effectiveness of the proposed method on the very well-known benchmark problem known as the Witsenhausen's counter-example. We note again that although our numerical results pertain to a particular problem, our approach is general and applicable to any problem of this nature (see e.g. \cite{basar2008variations} for some variations of Witsenhausen's counter-example). For example, when controllers have side information correlated with the source -i.e., when the problem setting resembles the classical Wyner-Ziv like problems in information theory- the cost surface becomes riddled with locally optimum points, see e.g. \cite{martins} for a control theoretic analysis of such a setting. The  greedy competitor approaches are expected to get  trapped in a local optimum while deterministic annealing is known to mitigate this problem, see \cite{da} for the effectiveness of DA on the similar optimization problems in quantization and clustering. Our future work will focus on such extensions.

\section*{Acknowledgments}
This work is supported by the NSF under grant  CCF 1118075. 


\bibliographystyle{IEEEbib_mod}

\bibliography{ref}\end{document}